\documentclass[5p, 10pt]{elsarticle}
\usepackage{times}
\usepackage{mathptmx}
\usepackage{amsmath,amssymb}
\usepackage[svgnames, dvipsnames, x11names]{xcolor}
\usepackage{lineno,hyperref}

\newcommand{\myvec}[1]{\mathbf{{#1}}}
\newcommand{\ignore}[1]{}

\begin{document}

\begin{frontmatter}

\title{Tether Capture of spacecraft at Neptune}

\author[adJRSL,adJRSLa]{J. R. Sanmart{\'{\i}}n}
\ead{juanr.sanmartin@upm.es}

\author[adJRSLa]{J. Pel{\'a}ez\corref{jpa}}
\cortext[jpa]{Corresponding author}
\ead{j.pelaez@upm.es}

\address[adJRSL]{Real Academia de Ingenier\'{\i}a of Spain}
\address[adJRSLa]{Universidad Polit{\'e}cnica de Madrid, Pz. C.Cisneros 3, Madrid 28040, Spain}

\begin{abstract}\fontsize{9}{10.8}\selectfont
Past planetary missions have been broad and detailed for Gas Giants, compared to flyby missions for Ice Giants. Presently, a mission to Neptune using electrodynamic tethers is under consideration due to the ability of tethers to provide free propulsion and power for orbital insertion as well as additional exploratory maneuvering --- providing more mission capability than a standard orbiter mission. Tether operation depends on plasma density and magnetic field $\mathbf{B}$, though tethers can deal with ill-defined density profiles, with the anodic segment self-adjusting to accommodate densities. Planetary magnetic fields are due to currents in some small volume inside the planet, magnetic-moment vector, and typically a dipole law approximation --- which describes the field outside. When compared with Saturn and Jupiter, the Neptunian magnetic structure is significantly more complex: the dipole is located below the equatorial plane, is highly offset from the planet center, and at large tilt with its rotation axis. Lorentz-drag work decreases quickly with distance, thus requiring spacecraft periapsis at capture close to the planet and allowing the large offset to make capture efficiency (spacecraft-to-tether mass ratio) well above a no-offset case. The S/C might optimally reach periapsis when crossing the meridian plane of the dipole, with the S/C facing it; this convenient synchronism is eased by Neptune rotating little during capture. Calculations yield maximum efficiency of approximately 12, whereas a $10^{\circ}$  meridian error would reduce efficiency by about 6{\%}. Efficiency results suggest new calculations should be made to fully include Neptunian rotation and consider detailed dipole and quadrupole corrections.
\end{abstract}


\end{frontmatter}


\section{Introduction}

Broad missions ---Cassini at Saturn, as well as Galileo and Juno at Jupiter--- have helped to provide significant knowledge about the Gas Giants. For minor missions involving specific visits such as exploring Jupiter's Europa or Saturn's Enceladus, electrodynamic tethers can provide free propulsion and power for orbital insertion, as well as further exploration \cite{sanmartin2005exploration} because of their thermodynamic properties --- and have the potential to create greater mission capabilities than a standard orbiter mission.

The two Ice Giants ---Uranus and Neptune--- have been considered by NASA as flagship missions for the next decade \cite{giants100520pre}. There are multiple points of interest in exploring Ice Giants that are different from those associated with Gas Giants:

\begin{description}
  \item[Exoplanet statistics] Data suggests that Ice Giants are much more abundant than Gas Giants.

  \item[Composition] Markedly different between Ice and Gas Giants ---Jupiter and Saturn are primarily made of hydrogen and helium--- while Uranus and Neptune contain substantial amounts of water, ammonia, and methane.

  \item[Dynamics] Shows singular features that are possible signs of collision with other big bodies in the intriguing, early Solar System dynamics. The rotation axis of Uranus nearly lies in the ecliptic plane itself. Neptune ---unlike the other Giants with many moons--- has just one large moon, Triton, which is in retrograde orbit.

  \item[Magnetic field structure] Neptune shows a striking difference that is highly important for tether interaction --- as detailed in the present analysis.
\end{description}

The following discusses whether tethers might be used for a minor mission to Neptune.  As in a Saturn case, tethers could appear inefficient as compared with Jupiter \cite{sanmartin2008electrodynamic}. In effect, the magnetic field $\mathbf{B}$ is similarly small, and spacecraft-capture efficiency (S/C-to-tether mass ratio, $M_{SC}/m_{t}$) decreases as ${B}^{2}$ for weaker fields. The S/C relative velocity  $ \mathbf{v}' \equiv \mathbf{v} - \mathbf{v}_{pl}$  induces in  the co-rotating magnetized ambient plasma a motional field $\mathbf{E}_{m} \equiv \mathbf{v}' \wedge \mathbf{B}$ in the S/C reference frame, and $\mathbf{B}$ exerts Lorentz force per unit length $\mathbf{I} \wedge \mathbf{B}$ on tether current $ \mathbf{I}$ driven by $\mathbf{E}_{m}$.

A disruptive bare-tether concept enhanced current flow by eliminating tether insulation and a large spherical conductor at the anodic end \cite{smmara}, making current and relative plasma-bias vary along the tether ---causing it to act as a giant Langmuir probe. Design capture-efficiency requirements for Jupiter are less than expected because its high $\mathbf{B}$ value might result in strong tether heating and/or energetic attracted electrons crossing the thin-tape tether and missing collection \cite{sanmartin2016analysis}. This requires limiting tether length $L_t$ –--to keep length-averaged current density well below a maximum: short-circuit value $\sigma_t \,E_m $ ($\sigma_t$ being tether conductivity)--- due to ohmic effects and proportional to field $B$. For weaker fields, tethers may avoid those issues, with current-density reaching  near the particular short-circuit maximum; capture-efficiency is approximately 3.5 for both Jupiter and Saturn \cite{sanmartin2018comparative}.

Section \ref{sec.2} introduces Neptune's environment issues concerning incomplete data on the ionospheric plasma electron density $N_{e}$ and magnetic field structure. Section \ref{sec.3} considers parameters of Neptune as a viable planetary tether environment, which is standard in an operation scheme. In Section \ref{sec.4}, a simple version of the complex dipole model OTD2 ---regularly used in the literature on Neptunian magnetics--- is used to determine capture-efficiency. Results exceeding the Jupiter/Saturn values are discussed in Section \ref{sec.5} and Conclusions are summarized in Section \ref{conclusions}.

\section{Neptune environment issues}\label{sec.2}

Tether operation depends on electron density $N_e$ ---appearing in tether current--- and magnetic field in both motional field and Lorentz force. The required and available information for both electron density and magnetic field have drastically different characteristics. The short 1989 Voyager 2 flyby, which came very close to Neptune, made getting models from data a difficult task.

\subsection{Plasma density issues}

Voyager 2 provided three types of data that yielded conflicting descriptions of plasma density at Neptune, two involving instruments on board, a third one using radio signals from the spacecraft to Earth. First, there are in-situ data from measurements by the PLS (plasma) instrument, using 4 modulated-grid Faraday cup detectors \cite{belcher1989plasma}. Next, a Plasma Wave instrument receiving wave (particularly whistler wave) data from a Planetary Radio Astronomy on-board antenna provided electric-field wave data into a wideband waveform receiver \cite{gurnett1990whistlers}. Finally, radio signals from Voyager 2 to Earth, following S/C occultations by Neptune, allowed tracking stations in Australia and Japan to determine ionospheric $N_e$  profiles versus altitude \cite{lindal1992atmosphere}.

The Faraday cups showed maximum density $N_e \sim 2$ cm$^{-3}$ during the brief closest approach at $79^{\circ}$ North, $ r \approx 1.18\ R_N$, or about 4,500 km altitude \cite{richardson1995plasma} --- while both radio-antenna and tracking-stations showed densities orders of magnitude above. Tracking data ranged from values of order $10^4$ cm$^{-3}$  around 1,250 km, down to $10^2$ cm$^{-3}$  in the 2,500/5,000 km altitude range \cite{bishop1995middle}. The waveform receiver showed intermediate densities ---around $10^3$ cm$^{-3}$--- at in-between altitudes \cite{gurnett1995plasma}.

\begin{figure}[!h]
\centerline{\includegraphics[width=65mm, clip]{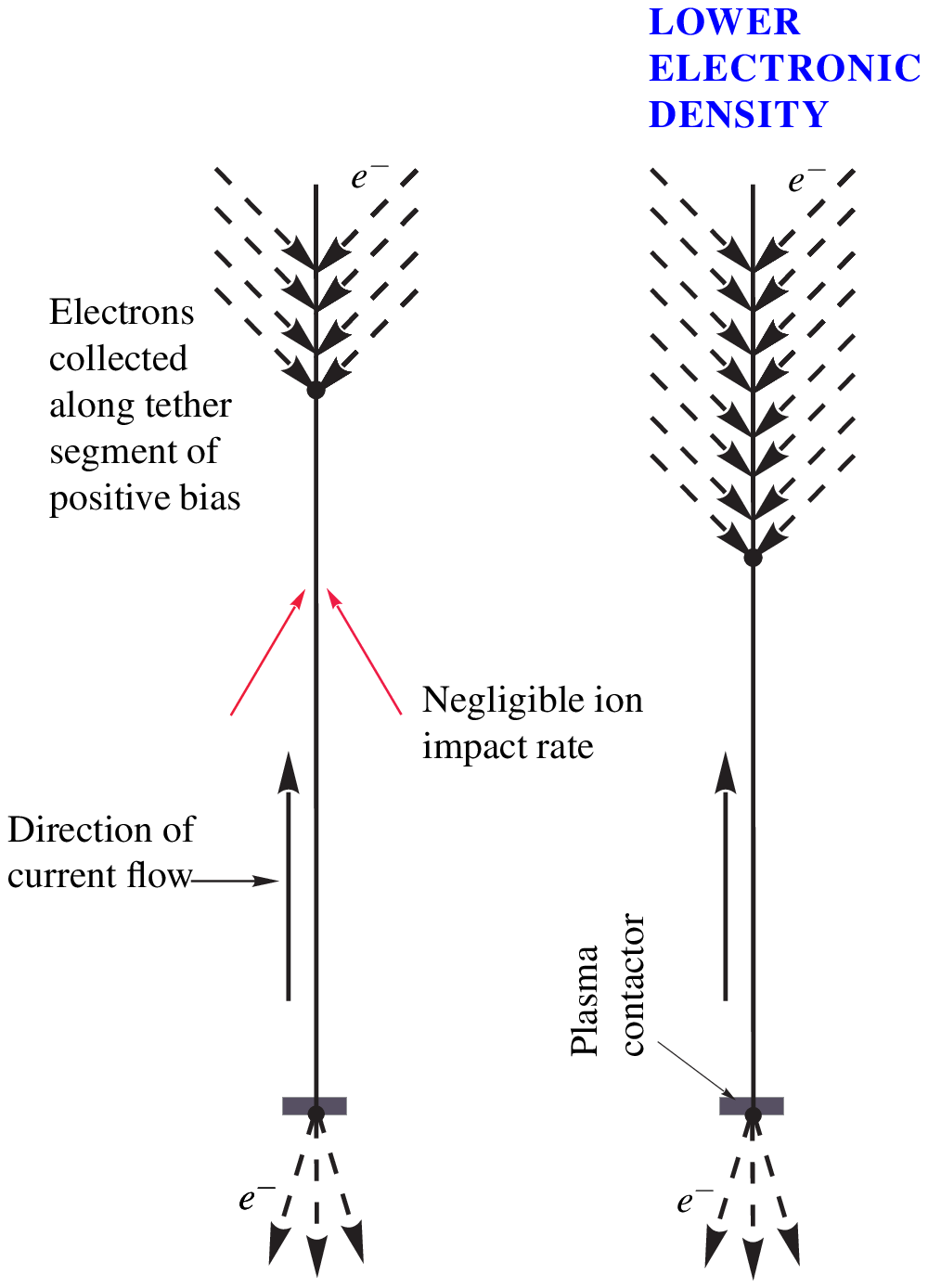}}
\caption{Interaction tether-ambient plasma} \label{fig1}
\end{figure}

Modeling of whistler propagation did present some issues \cite{gurnett1995plasma}. More importantly, regarding conflicting data, the question was raised whether Voyager 2 had entered Neptune's ionosphere --- or whether a cold-density plasma component was detectable by the PLS instrument, which might be interacting with precipitating particles and not co-rotating flow \cite{richardson1995plasma}. Further, radio-tracking data above 2,500 km were problematic because of variations in the ionosphere and interplanetary medium that the radio signal traverses \cite{bishop1995middle}.

Certainly, data from the Voyager 2 flyby did not yield a full model of the ambient magnetized plasma around Neptune. Tether operation, however, would first involve plasma density in the lower, no-conflict ionosphere, and secondly deal with ill-defined density profiles by appealing appropriately sized tapes. A bare tether accommodates density variations by allowing the anodic segment to self-adjust by varying its fraction of tether length (Figure 1).   This could make a broad range of $N_e$ values lead to length-averaged current near the short-circuit maximum $w_t\,h_t \times \sigma_t\,E_m$ (a  design reference-value independent of actual electron densities), for a convenient range of $L_t$ values, not affecting tether mass if tether-tape width $w_t$ is adjusted at a fixed thickness $h_t$.

\begin{figure}[!h]
\centerline{\includegraphics[width=80mm, clip]{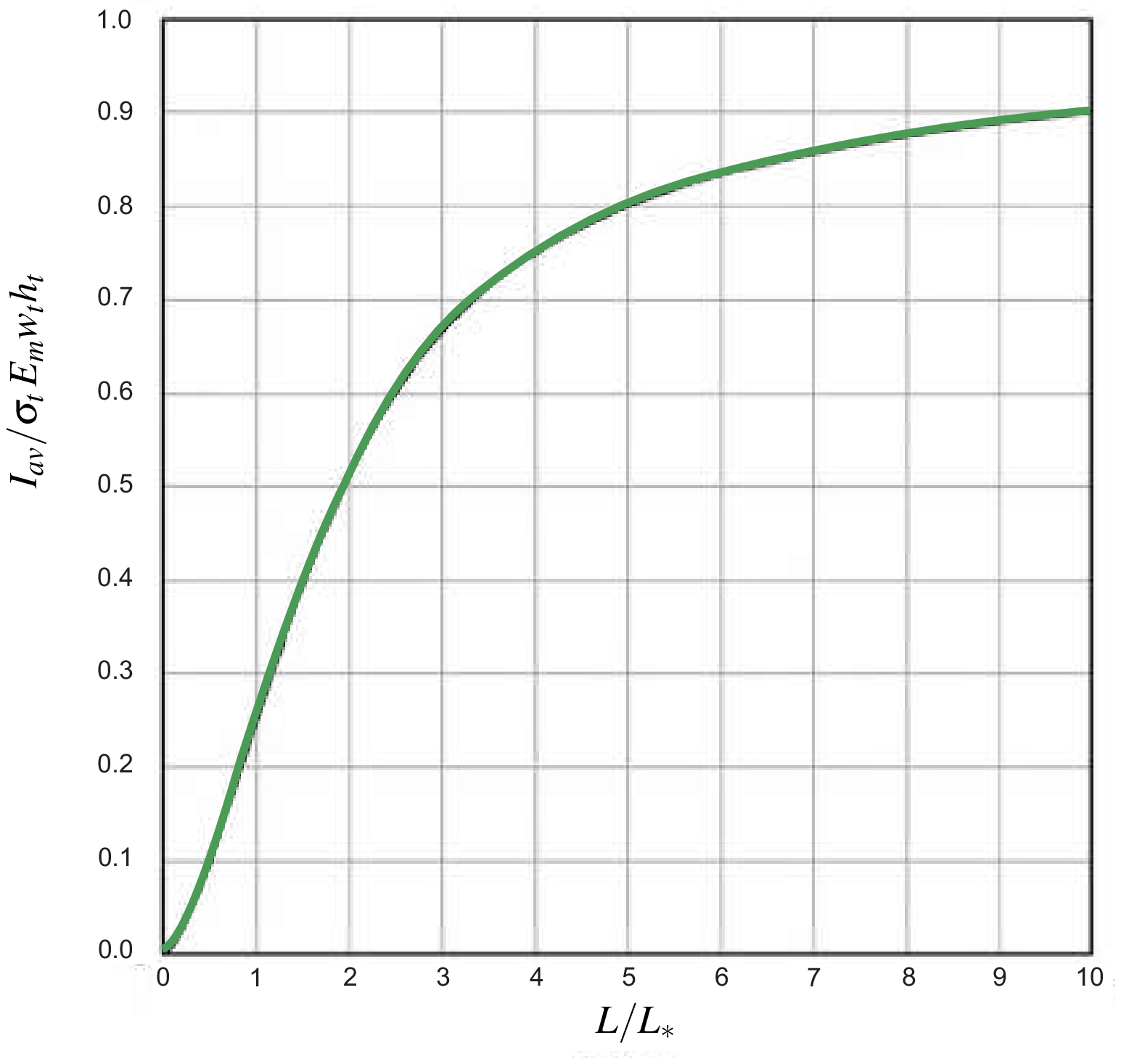}}
\caption{Length-averaged current vs. tether length} \label{fig2}
\end{figure}

In the OML (orbital-motion-limited) electron-collection regimen of bare tethers, the Lorentz force on the tether is proportional to its length-averaged current,
\begin{align}
 I_{av} &= w_{t}h_{t} \sigma_{t} E_{m} \times i_{av} ( L_{t} / L_{\star}) \label{eq.1} \\
  L_{\star} &\propto E_{m}^{1/3}(\sigma_{t}\,h_{t} /N_{e})^{2/3}         \label{eq.2}
\end{align}
where $L_{\star}$  is a characteristic length gauging the bare-tether electron-collection impedance against ohmic resistance and increasing with decreasing $N_{e}$  \cite{sanmartin2008electrodynamic}. The dimensionless average-current $i_{av}< 1$  (see Figure  \ref{fig2}), which vanishes with length ratio $L_{t} / L_{\star}$, approaches 1 at large $L_{t} / L_{\star}$ following the law:
\begin{equation}\label{eq.3}
  i_{av} = 1 - L_{\star}/ L_{t} \quad \mbox{for} \quad L_{t} / L_{\star} > 4.
\end{equation}
Equation \eqref{eq.3} shows effective OML-current accommodation to drops in electron density. Consider $L_{t} / L_{\star}$  decreasing from 20 to 4, with tether length $L_{t}$ and parameters in equation \eqref{eq.2} other than $N_{e}$ to be constant.  The value of $N_{e}$ itself would then be lower by a 0.09 factor, whereas current $i_{av}$ in \eqref{eq.3} would decrease from 0.95 to 0.75.

\subsection{Magnetic field issues}

Regarding magnetic fields, $\myvec{B}$, the fields of Gas Giants are present due to currents from charges repeatedly moving in some small volume deep inside the planet. Outside the planet, at ``large" distances from the system of steady currents but close to the planet, the field is generally described through the universal magnetic-moment vector concept and its dipole-law approximation.  A magnetic moment $\myvec{m}= m\, \myvec{u}_{m}$ of magnitude $m$ ({gauss}$\times${meter}$^{3}$) and unit vector $\myvec{u}_{m}$ located at a ``point" $\myvec{r}_{m}$  gives a definite magnetic field at a ``faraway" point $\myvec{r}$ \cite{landau1962classical}
\begin{equation}\label{eq.4}
\myvec{B}(\myvec{r}) \approx \left [ 3(\myvec{u}_{\rho} \cdot \myvec{m})\, \myvec{u}_{\rho} - \myvec{m} \right ]/ \rho^{3}
\end{equation}
with $\myvec{\pmb{\rho}} = \rho\, \myvec{u}_{\rho} \equiv \myvec{r} - \myvec{r}_m$ vector position in planetary
frame with displaced origin. For Saturn, $\myvec{m}$ is at its center (${r}_{m} \approx 0$, $\myvec{\pmb{\rho}} = \myvec{r}$)
and near parallel to its rotation axis --- with Jupiter also having relatively similar conditions. At points in a circular equatorial orbit, the field would then read $\myvec{B}(\myvec{r}) = - \myvec{m}/r^3$,  and the Lorentz force on a tether, orbiting vertical, is parallel to the velocity.

For Neptune, a dipole law with $\myvec{r}_{m}$  well off-center and $\myvec{u}_{m}$  complex  orientation is valid for somewhat large $r/R_{N}$  \cite{ness1989magnetic} distances that are of no interest for tether applications due to the Lorentz force, which is quadratic in the field, decreases with the inverse $6^{th}$ power of distance to the planet.  In the OTD2 offset-tilted detailed model  \cite{connerney1991magnetic}, with magnetic moment of magnitude $m = 0.13\  \mbox{G}\times R_{N}^{3}$, the dipole is  located $0.19\ R_{N}$  below the equatorial plane and $0.52\ R_{N}$  radially away from the rotation axis ---in the plane containing axis and dipole--- where $\myvec{u}_{m}$ is tilted $47^{\circ}$ with respect to the axis and $22^{\circ}$ off the meridian plane.

For $r < 2.5\ R_{N}$, quadrupole terms in a multipole expansion of field $\myvec{B}$ decrease faster than the dipole ---and/or additional multipoles from differently localized current sources--- and may generally need be considered at difference with Jupiter and Saturn \cite{ness1995neptune, connerney1993magnetic,russell2010magnetic}. Alternative spherical harmonic analysis has reached a reasonable quadrupole description, though relevance is dependent on longitude and latitude ranges. For simplicity in the present analysis we shall keep the above dipole term with a reduced approximation.

\section{Neptune particular S/C-capture regime}\label{sec.3}

For the hyperbolic orbit of a S/C in Hohmann transfer between heliocentric circular orbits at Earth and Neptune, the arriving velocity in the Neptune frame is about $v_{\infty} = 3.96$~km/s, the orbital specific energy $\varepsilon_{h}$ being $v_{\infty }^{2}/2$. Using that value in the general relation between $\varepsilon$ and eccentricity $e$ at given periapsis $r_{p}$,
\begin{equation}\label{eq.5a}
  \varepsilon =  \mu_{N} (e - 1)/ (2 r_{p})
\end{equation}
yields a hyperbolic eccentricity very close to unity,
\begin{equation}\label{eq.5b}
  e_h - 1 = v_{\infty}^2\ R_N / \mu_N = 0.058
\end{equation}
where $\mu_{N}\approx  6835107$ km$^{3}$/s$^{2}$ is the gravitational constant, and we conveniently set the periapsis very close to the planet (as in Jupiter and Saturn analyses \cite{sanmartin2016analysis,sanmartin2018comparative}) $r_{p}\approx R_{N} \approx $ 24,765 km. Lorentz-drag capture results in barely elliptical orbits (eccentricity $e_{c}$ just below 1). For conditions of interest, calculations will be carried out using a parabolic orbit throughout, with no sensible change except moving from open to closed. The required capture-dynamics is, in a sense, weak.

When the relative velocity $\myvec{v}'$ opposes $\myvec{v}$ the Lorentz force is actually thrust. This is the case for prograde circular orbits beyond a radius $a_{s}$ where plasma co-rotating with planet spin and orbital velocities are equal.
\begin{equation}\label{eq.6}
 \Omega\, a_S = \sqrt{\mu/a_S} \quad \Rightarrow \quad a_S/R \propto \left( \rho_{pl}/\Omega^2 \right )^{1/3}
\end{equation}
Among Giants, Neptune and Saturn have the highest and lowest densities, respectively, while both Ice Giants have spin slower than Gas Giants. This results in values $a_{s}/R_S \approx  1.89$ and $a_{s}/R_N \approx 3.41$, with corresponding values ---compared to Jupiter and Earth--- in between the two, and larger than either, respectively. For the parabolic orbits of interest, the radius where drag vanishes along with the relative tangential velocity $v'_{t}$ is \cite{sanmartin2008electrodynamic}
\begin{equation}\label{eq.7}
r_M/R = \sqrt{2} \left( a_S/R \right )^{3/2} \propto \left( \rho_{pl}/\Omega^2 \right )^{1/2}
\end{equation}
varying from 3.7 for Saturn to 8.9 for Neptune.

\begin{figure}[!h]
\centerline{\includegraphics[width=80mm, clip]{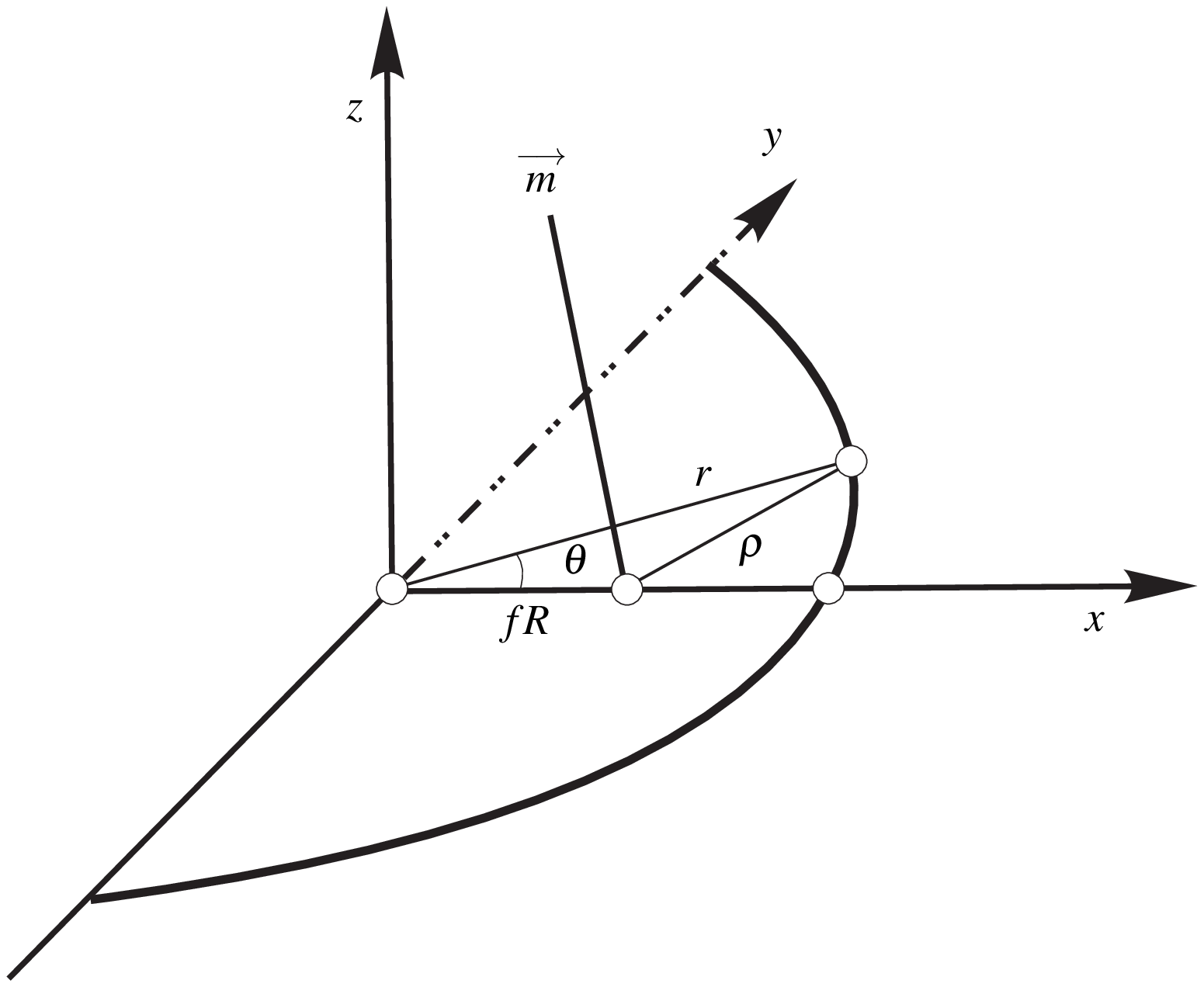}}
\caption{Magnetic dipole and \textit{S/C} trajectory} \label{fig3}
\end{figure}

This makes for important capture differences at Neptune, with most of the drag arc from $r_{p}$ to $r_{M} \gg r_{p}$ being irrelevant, because Lorentz drag becomes negligible well before $r$ reaching $r_{M}$, quite opposite the case with Saturn, which made its retrograde capture convenient \cite{sanmartin2018comparative}.  This is manifest here in the velocity at the parabolic periapsis, $v_{p} = \left( 2\mu_{N}/R_N\right)^{1/2}  \approx $ 23.5~km/s being much larger than co-rotation velocity $2\pi R_{N}$ /16.1 h $\approx $ 2.7~km/s, and allows using $\myvec{v}' \approx \myvec{v}$  throughout the relevant drag arc.

Furthermore, because of the large dipole offset ---equal moment $m$ in capture is greatly more efficient than for a no-offset case--- the S/C might optimally reach periapsis when crossing the meridian plane that contains the dipole center, resulting in the S/C facing the dipole when at periapsis (Figure  \ref{fig3}). This convenient synchronism is somewhat eased by Neptune having slow spin and high density ---with the high density making S/C orbital motion fast--- as following from the Barker equation for parabolic orbits, giving time $t$ from periapsis-pass versus radius,
\begin{equation}\label{eq.8}
  (3\,v_p/2r_p)\,t = \sqrt{\tilde{r}-1}\ (2+ \tilde{r}), \qquad (\tilde{r}\equiv r/r_p)
\end{equation}
The characteristic time of the S/C motion, for $r_{p}  \approx R$, is thus of order
\begin{equation}\label{eq.9}
r_{p} /v_{p} \approx \sqrt {R^{3}/2GM_{pl}}\ \propto \ 1/\sqrt{\rho_{pl}}.
\end{equation}
Because of the slow planet spin and the short range of Lorentz contribution to its drag work, we will here estimate  $\rho $ distances while ignoring the rotation of Neptune and its $\myvec{r}_{m}$  position.

\section{Capture-efficiency calculation}\label{sec.4}

\begin{figure}[!h]
\centerline{\includegraphics[width=80mm, clip]{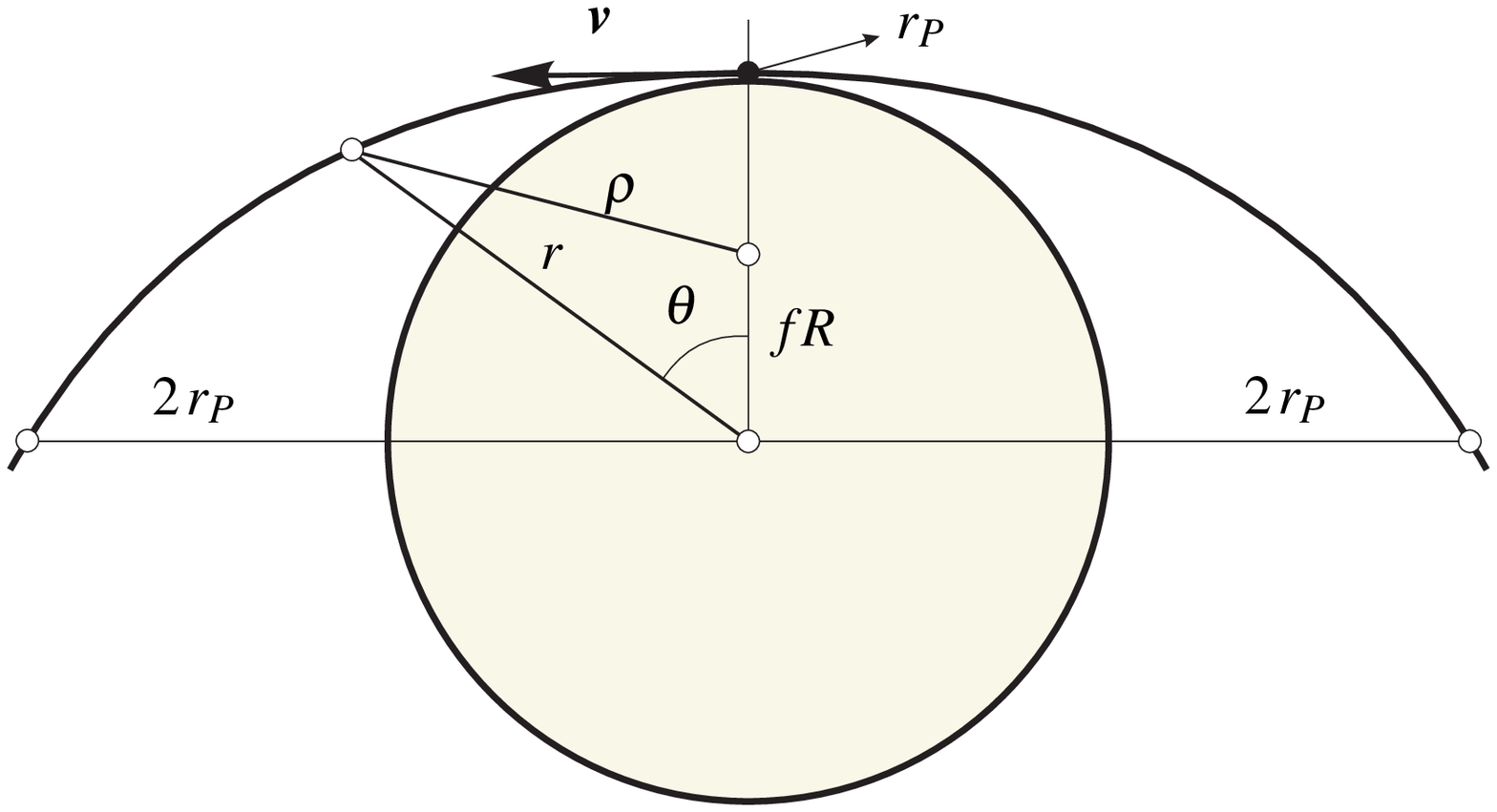}}
\caption{Full synchronism case} \label{fig4}
\end{figure}

In the present estimate, we ignore both distance $0.19\ R_{N}$ and the $22^{\circ}$ angle. We consider Neptune magnetics as presenting a $0.55\ R_{N}$ offset and a $47^{\circ}$  tilt, both having a substantial effect on capture efficiency. Also, as in \cite{sanmartin2008electrodynamic,sanmartin2016analysis,sanmartin2018comparative}, we shall consider a S/C approaching Neptune in an equatorial orbit. Writing $\myvec{B}\,=\,\myvec{B}_{ax} + \myvec{B}_{eq}$, where subscripts \textit{ax} and \textit{eq} stand for field components along the Neptunian rotational axis and in the equatorial plane, respectively, we may rewrite a standard power law $\dot{{W}}_{L}
\,\,=\,\,\myvec{{v}}\,\cdot \,(L_{t} \myvec{{I}}_{av} \wedge \myvec{B})$ as
\begin{equation}\label{eq.10}
 \dot{W}_L = \myvec{v}\cdot (L_{t} \myvec{{I}}_{av} \wedge \myvec{B}_{ax}) = - L_t \myvec{I}_{av} \cdot (\myvec{v} \wedge \myvec{B}_{ax})
\end{equation}
All three vectors $\myvec{v}$, $\myvec{I}_{av}$ and $\myvec{B}_{eq}$ lie in the equatorial plane, thus having no joint power effect (Figures \ref{fig3},\ref{fig4}). Having ignored the $0.19\ R_{N}$   distance parallel to the Neptunian axis makes the second term in Equation \eqref{eq.4} contribute to $\myvec{B}_{ax}$.

\subsection{Full synchronism case}

\begin{figure}[!h]
\centerline{\includegraphics[width=80mm, clip]{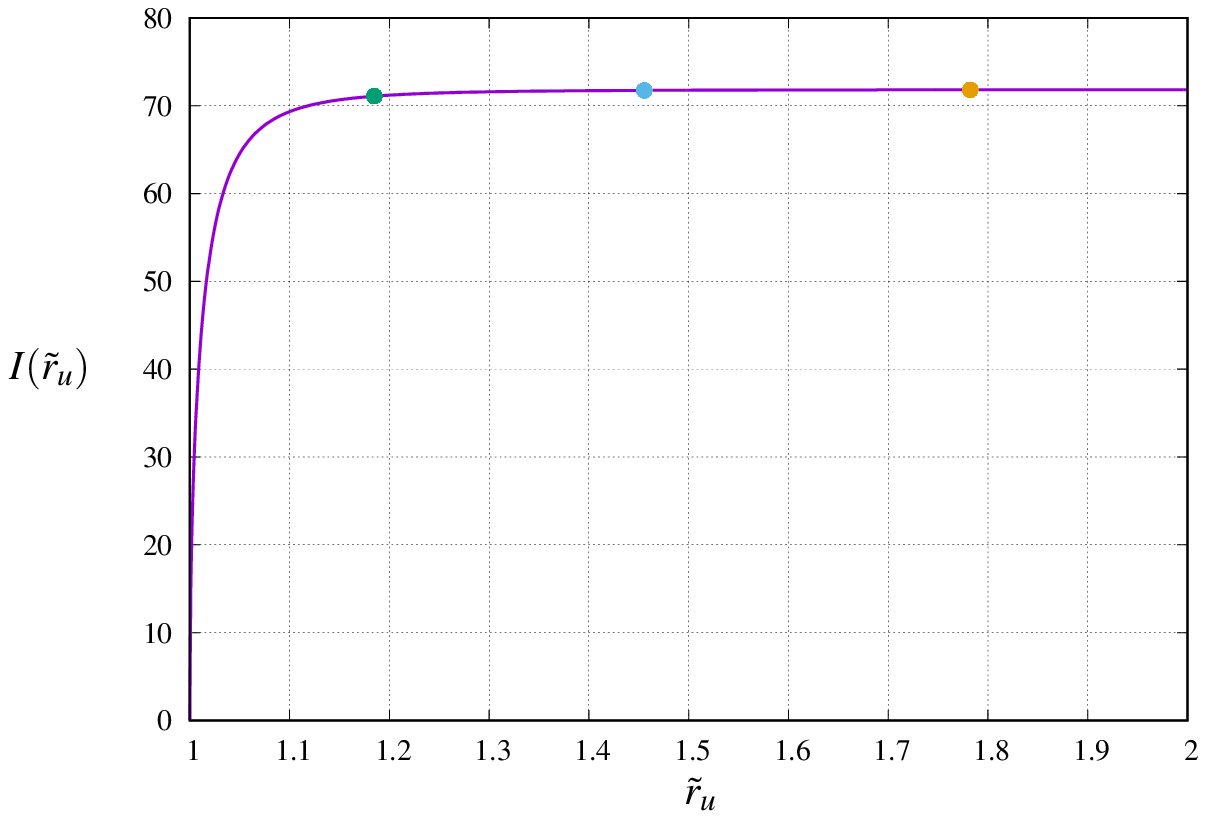}}
\caption{$I(\tilde{r}_u)$. Points correspond to $n=2,3,4 $} \label{fig5}
\end{figure}

Using Equation \eqref{eq.1} with $m$-subscript omitted, we write
\begin{equation}\label{eq.11}
L_{t} \,\myvec{I}_{av} = L_{t} w_{t} h_{t} \sigma_{t} i_{av} \times E_{eq} \myvec{u}_{I}
\end{equation}
where unit vectors along tether $(\myvec{u}_{I})$ and $\myvec{E}_{eq}$ are
both radial at periapsis and nearly so in a short effective-drag arc in
 Figure \ref{fig4}. Also, using $\myvec{v}' \approx \myvec{{v}}$ to write $E_{eq}
\myvec{u}_{I} \approx \myvec{E}_{eq} \approx \myvec{v} \wedge \myvec{B}_{ax} $ in Equation \eqref{eq.10}, Lorentz power becomes
\begin{equation}\label{eq.12}
\dot{W}_{L} (r,\rho) = -\dfrac{m_{t}}{\rho_{t}}\,\sigma_{t} i_{av} \times \,v^{2}\,B_{ax}^{2} =-\dfrac{m_{t}}{\rho_{t}}\,\sigma_{t} i_{av} v^{2}\left (m\times \cos 47^{\circ}/\rho^{3}\right)^{2}
\end{equation}
At full synchronism, the symmetry relation between $\rho $ and $r$ at positive
and negative values of true anomaly $\theta$ leads to equal Lorentz-drag work in
symmetric arcs at periapsis. We may thus write the Lorentz-drag work as
\begin{equation}\label{eq.13}
W_{L}  = \int_{\Delta t} \dot{W}_{L}\,  \mathrm{d}t = 2\, \times\,\int_{R_{N}}^{r_{u}} {\dfrac{\dot{W}_{L} \mathrm{d}r}{\mathrm{d}r/\mathrm{d}t}}
\end{equation}
the upper limit $r_{u}$, though well below $r_{M}$, is here considered ---to include the radial arc--- and contributes to drag. We then have
\begin{equation}\label{eqz.1}
 W_{L} \,=\,-\,\dfrac{m_{t}}{\rho_{t}}\,\sigma_{t} \,\left(0.13\,GR_{N}^{3} \,\cos 47^{\circ}\right)^{2}\,\left\langle {i_{av}}\right\rangle_{r} \,\int_{R_{N}}^{r_{u}} \dfrac{v^{2}}{\rho
^{6}}\,\dfrac{\,2\times r \mathrm{d}r}{\sqrt {2\mu_{N} (r\,-R_{N})}}
\end{equation}
where we used the radial speed-rate in parabolic arcs, $\mathrm{d}r/\mathrm{d}t = [2\mu_N (r -R_N)]^{1/2}/r$, finally yielding
\begin{align}
 W_{L} &=-\dfrac{m_{t}}{\rho_{t}}\,\sigma_{t} \left(0.13\,G\,\cos 47^{\circ}\right)^{2} \,\left\langle {i_{av}}\right\rangle_{r} v_{p} R_{N} \times I(\tilde{{r}}_{u}) \label{eq.14a} \\
 I(\tilde{{r}}_{u} ) &\equiv \int_1^{\tilde{{r}}_{u}} \dfrac{\,2\times \mathrm{d}\,\tilde{r}}{\tilde{{\rho}}^{6}\sqrt{\tilde{r}-1}} \label{eq.14b}
\end{align}
$\tilde{\rho}$ and $\tilde{r}$ keeping a simple relation along the S/C orbital motion (see Figure \ref{fig4}).
\begin{equation}\label{eq.15}
\tilde{\rho}^{2} = \left(\tilde{r} + f \right)^{2} -4f, \quad (\tilde{\rho}\,\equiv \,\rho/R_N,\quad \tilde{r}\,\equiv \,r/R_N,\quad f=0.55).
\end{equation}
Using Equation (\ref{eq.5a}-\ref{eq.5b}) in drag-work per unit mass
\begin{equation}\label{eqz.2}
\dfrac{\left| {W_{L}} \right|}{M_{SC}} = \varepsilon_{h} - \varepsilon_{c} = \dfrac{\mu_N}{2\,r_{p}} \left( e_{h} - e_{c}\right) =  \dfrac{v_{\infty }^{2}}{2}\,\dfrac{e_{h} - e_{c}}{e_{h} -1},
\end{equation}
and Equation \eqref{eq.14a} then yields capture efficiency
\begin{align}
 \dfrac{M_{SC}}{m_{t}}\,&<\, \dfrac{M_{SC}}{m_{t}}\,\times \,\dfrac{e_{h} -e_{c}}{e_{h}-1} \,=\,\dfrac{\left|W_{L} \right|}{m_{t} v_{\infty}^{2}/2}\,=    \notag \\
 &=\,2\,\dfrac{\sigma_{t} v_{p} R_{N}}{\rho_{t} v_{\infty}^{2}}\,\left( 0.13\,G\,\times \,0.68 \right)^{2}\,\left\langle
i_{av} \right\rangle_{r} \,\times \, I(\tilde{{r}}_{u})  \label{eq.16}
\end{align}
At the lower limit in the integral, $\tilde{r}_{l} \,=\,1$, Equation \eqref{eq.15}
gives $\tilde{\rho}_{l} =1-f = 0.45$. Now we consider upper values, $\tilde{\rho}_{u}
=n (1-f)\,\equiv \,\tilde{\rho}_{n}$, with $ n> 1 $ yielding for
$\tilde{r}_{u}$, at Equation \eqref{eq.15},
\begin{equation}\label{eq.17}
\tilde{r}_{u} = \tilde{r}_{n} \,\equiv \,\sqrt{4f + \tilde{\rho}_{n}^{2}} -f, \qquad n > 1
\end{equation}
The integral $I(\tilde{{r}}_{u})$, ---which is not singular as moving to a variable $\tilde{r} \equiv  1 + z^{2}$ would manifest--- converges rapidly, as shown in  Table \ref{t1} and Figure \ref{fig5}, allowing integration to stop at $r \approx  1.12\,r_{p}$ in the parabolic orbit, with a
value around 70.1.

\begin{table}
\begin{center}
\begin{tabular}{|l|l|l|l|}  \hline
 $n$ & $\tilde{r}_u$ & $\tilde{\rho}_u$ & $I(\tilde{r}_u)$ \\ \hline
 1.5 & 1.080 & 0.675 & 68.16  \\ \hline
 1.7 & 1.120 & 0.765 & 70.10 \\ \hline
 2   & 1.185 & 0.90 & 71.088 \\ \hline
 3   & 1.456 & 1.35 & 71.752 \\ \hline
 4   & 1.782 & 1.80 & 71.818 \\ \hline
\end{tabular}
\end{center}
\caption{$I(\tilde{r}_u)$ for different values of $n$} \label{t1}
\end{table}

For an aluminum tether, Equation \eqref{eq.16} then provides maximum capture-efficiency
$M_{SC}/m_{t} = 5.35 \times \langle i_{av} \rangle_{r}$. Similar to the Saturn research \cite{sanmartin2018comparative}, efficiency can be increased by a gravity assist from Jupiter through a tail flyby, increasing the heliocentric S/C velocity at the encounter with Neptune ---which would be greater than the velocity from a direct Hohmann transfer. This reduces the hyperbolic velocity relative to the planet by a 1.53 factor, from 3.96 to 2.59~km/s. Capture efficiency then increases in Equation \eqref{eq.16} by a 2.34 factor to $12.52 \times \langle i_{av} \rangle_{r}$.

Regarding the average dimensionless current $i_{av}(L/L_{\star})$,
the expression for $L_{\star}$ in Equation \eqref{eq.1} reads, for aluminum \cite{ahedo2002analysis},{\fontsize{9}{10.8}\selectfont
\begin{align}
  L_{\star} &\equiv \,\dfrac{(m_{e} E_{m})^{1/3}}{2^{7/3}e}\,\left(
3\pi \dfrac{\sigma_{t} h_{t}}{N_{e}} \right)^{2/3} \label{eqz.3} \\
   L_{\star} & \approx \,2.78\, \mbox{km}\, \cdot \,\left( {\dfrac{E_{m} }{100\, \mbox{V/km}}} \right)^{1/3}\,\cdot
\,\left( {\dfrac{h_{t} }{100\,\mu\mbox{m}}\,\cdot \,\frac{10^{5}/\mbox{cm}^{3}}{N_{e}}}
\right)^{2/3} \label{eq.18}
\end{align}}\relax
Because the drag-arc is short we set $E_{m}(r) \approx E_{mp}$ throughout
the $r$-integration. Using
\begin{equation}\label{eqz.4}
 B_{p} = \dfrac{0.68 \times 0.13}{0.45^{3}} \times (10^{-4}\, \mbox{Vs/m}^{2}) = 0.97 \times 10^{-4}\, \mbox{Vs/m}^{2}
\end{equation}
we find that $E_{mp} = v_{p} B_{p} = 22.7 \times 100$ V/km. Setting
$h_{t} = 10\,\mu\mbox{m}$, $N_{e} = 10^{3}$ cm$^{-3}$, yields $L_{\star} =36.6$~km. A tether length $L =73.2$~km leads to $i_{av}(2) = 0.52$ and to an efficiency 6.51 in the case of a Jupiter flyby. For a density of $10^{4}$ cm$^{-3}$, $L_{\star}$ would decrease by a factor of 0.215, and $L/L_{\star}$,
$i_{av}$ and capture efficiency would increase to values 9.30, 0.9, and 11.3, respectively.

\subsection{Synchronism mismatch}

If the S/C reaches periapsis before or after crossing the meridian plane containing the dipole, Equation \eqref{eq.13} is no longer valid, with capture drag-work being the sum over drag work for positive (negative) \textit{true-anomaly} $\theta$ ranges, which are differently drag-effective. Full work is the same, however, whether the dipole lies at $\theta$ positive or negative; for definiteness, set the dipole at a positive value  $\alpha$.  Equation \eqref{eq.14b} takes now the form
\begin{align}
 I(\tilde{r}_{u})\,&\equiv\, \int_1^{\tilde{r}_{u}} \dfrac{\mathrm{d} \tilde{r}}{\tilde{\rho}_{-}^{6} \sqrt{\tilde{r} -1}} + \int_1^{\tilde{r}_{u}} \dfrac{\mathrm{d}\tilde{r}}{\tilde{\rho}_{+}^{6} \sqrt {\tilde{r}-1}} = \notag \\
 &=\int_1^{\tilde{r}_{u}} \dfrac{\mathrm{d}\tilde{r}}{\tilde{\rho}_{-}^{6} \sqrt {\tilde{r}-1}} \left( 1 + \dfrac{\tilde{\rho}_{-}^{6}}{\tilde{\rho}_{+}^{6}} \right) \label{eq.19}
\end{align}
where
\begin{align}
\tilde{\rho}_{\pm}^{2} &= \left( \tilde{r} + f\right)^{2} - 4f + 2f\,\left[ \left( 2-\tilde{r}\right) \left( 1- \cos\alpha \right) \pm 2\,\sqrt{\tilde{r} - 1} \sin \alpha \right] \notag \\
&(\tilde{\rho}\,\,\equiv \,\,\rho/R_N,\,\,\,\,\,\tilde{r}\,\,\equiv \,\,\,r/R_N)  \label{eq.20}
\end{align}
$+$ (-) signs corresponding to distances from the dipole to points in the parabolic orbit of common $r$ value and negative (positive) true anomaly. The parenthesis inside the integral starts at a value of 2 at the lower limit $\tilde{r}_{l} =1$, and decreases fast as $\tilde{r}$ increases.

At that lower limit $\tilde{r}_{l} = 1$ Equation  \eqref{eq.20} gives a common value,
\begin{equation}\label{eqz.5}
\tilde{\rho}_{l} = (1 - f)\,\sqrt{1 + \dfrac{4\,f\sin^{2}(\alpha/2)}{(1-f)^{2}}}
\end{equation}
Let us now consider upper limits for the integral in \eqref{eq.19},
$\tilde{r}_{u} = \tilde{r}_{n} $, corresponding in \eqref{eq.20}
to $\tilde{\rho}_{-} \equiv n\,\tilde{\rho}_{l} \ n> 1$. Results for the work
integrals are given in Table \ref{t2}, showing that capture-efficiency decrease would be about 1.8 {\%} for $\alpha = 5^{\circ}$. Decrease is 6{\%} for $\alpha = 10^{\circ} $.

\begin{table}[!h]
\begin{center}
\begin{tabular}{|l|l|l|l|l|}  \hline
 $n$ & $\tilde{r}_u$ & $\tilde{\rho}_u$ & $I(\tilde{r}_u)$ & $\frac{M_{SC}}{m_t}$ decrease \\ \hline
 1.5 & 1.100 & 0.682 & 68.11 &1.8 \%  \\ \hline
 1.7 & 1.144 & 0.774 & 69.42 &1.7 \% \\ \hline
 2   & 1.214 & 0.909 & 70.19 & 1.6\% \\ \hline
\end{tabular}
\end{center}
\caption{$I(\tilde{r}_u)$ for different values of $n$. Small mismatch $\alpha= 5^{\circ}$} \label{t2}
\end{table}

\section{Discussion of Results}\label{sec.5}

The result of our analysis is the determination of capture efficiency, defined as a ratio between mass $M_{SC}$ of captured S/C and mass $m_{t}$ of tether required for the capture (included in $M_{SC}$). We found that the large offset of the Neptune dipole resulted in it definitively exceeding the Saturn capture efficiency, thus compensating for the negative effect of the large tilt and the low value of magnetic moment 0.13 $G\,R_{N}^{3}$ against 0.21~$G\,R_{S}^{3}$. As in the Saturn case \cite{sanmartin2018comparative}, a gravity assist from Jupiter helped an efficient capture.

High efficiency in the presence of a large dipole-offset would require the S/C to arrive at periapsis ---at a point very close to the planet---  lying in the meridian plane of the Neptune dipole.  That synchronism was found to be reasonably necessary, whether the S/C reached that plane following, or ahead of, arrival at periapsis, a $10^{\circ}$ mismatch resulted in  a 6 {\%} decrease in efficiency.

Our calculations used several simplifications. One was ignoring the planet rotation while the S/C moved over the quite limited parabolic arc contributing to drag. This arose from distinctive properties of Neptune among the 4 Giant Planets. First, along with Uranus, it has a slow spin as compared with Gas Giants. Secondly, it has the highest density; for a parabolic orbit with periapsis close to a planet, the characteristic time of S/C motion is shorter for higher densities.

Another simplification involved the  dipole model used. The  dipole  lies $0.19\, R_{N}$  below the equatorial plane; we ignored this fact. Only the component of field $\myvec{B}$ along the rotational axis contributes to drag in an equatorial orbit. Our approximation excluded the first term in Equation \eqref{eq.4} from contributing to drag, because the dipole and S/C laid in the equatorial plane. With the dipole below the equatorial plane, the first term in Equation \eqref{eq.4} would contribute to drag.

Furthermore, quadrupole terms, which are more important than in Jupiter and Saturn cases because of the proximity of the magnetic moment to the planet surface, were fully ignored. Spherical harmonic analysis first provided roughly resolved harmonic coefficients describing the quadrupole \cite{connerney1991magnetic}, then reached a consistent description \cite{ness1995neptune}, \cite{connerney1993magnetic}.

\section{Conclusions} \label{conclusions}

The high capture efficiency results suggest new calculations to include Neptune rotation and to use the OTD2 detailed dipole-model, considering further quadrupole corrections; consideration of planet rotation would require revising the sensitivity of capture efficiency to synchronism mismatch. The upgraded analysis would allow the start of planning a visit to moon Triton mission.

\section{Competing Interests Statement}

There are no competing interests amongst the authors.

\section{Funding}

This work was supported by MINECO/AEI and FEDER/EU under Project ESP2017-87271-P.

\section{Acknowledgements }

The authors thank the MINECO/AEI of Spain for their financial support.
Part of J. Peláez's contribution was made at the University of Colorado
at Boulder on a stay funded by MICINN (ref. PRX19/00470) and
the Fulbright Commission, to which he thanks the support received.

\ignore{
\section*{References}
\bibliographystyle{elsarticle-num}
 \bibliography{neptuno}
 }

\end{document}